\documentclass[10pt,twocolumn,secnumarabic,amssymb, nobibnotes, aps, prd,superscriptaddress]{revtex4-2}
\usepackage{xcolor,soul}
\sethlcolor{red}
\usepackage{amsmath}
\usepackage{wrapfig}
\usepackage[utf8]{inputenc}

\usepackage{lipsum}
\usepackage{subfigure}
\usepackage{wrapfig}
\usepackage{xparse,xcoffins}

\ExplSyntaxOn
\NewCoffin\imagecoffin
\NewCoffin\labelcoffin

\keys_define:nn { miguel/label }
 {
  label   .tl_set:N = \l_miguel_label_tl,
  labelbox .bool_set:N = \l_miguel_label_box_bool,
  labelbox .default:n = true,
  fontsize .tl_set:N = \l_miguel_label_size_tl,
  fontsize .initial:n = \footnotesize,
  pos .choice:,
  pos/nw .code:n = \tl_set:Nn \l_miguel_label_pos_tl { left,up },
  pos/ne .code:n = \tl_set:Nn \l_miguel_label_pos_tl { right,up },
  pos/sw .code:n = \tl_set:Nn \l_miguel_label_pos_tl { left,down },
  pos/se .code:n = \tl_set:Nn \l_miguel_label_pos_tl { right,down },
  pos/n .code:n = \tl_set:Nn \l_miguel_label_pos_tl { hc,up },
  pos/w .code:n = \tl_set:Nn \l_miguel_label_pos_tl { left,vc },
  pos/s .code:n = \tl_set:Nn \l_miguel_label_pos_tl { hc,down },
  pos/e .code:n = \tl_set:Nn \l_miguel_label_pos_tl { right,vc },
  pos .initial:n = nw,
  unknown .code:n   = \clist_put_right:Nx \l_miguel_label_clist
                       { \l_keys_key_tl = \exp_not:n { #1 } }
 }
\clist_new:N \l_miguel_label_clist
\box_new:N \l_miguel_label_box
\box_new:N \l_miguel_label_image_box

\NewDocumentCommand{\xincludegraphics}{O{}m}
 {
  \group_begin:
  \tl_clear:N \l_miguel_label_tl
  \clist_clear:N \l_miguel_label_clist
  \keys_set:nn { miguel/label } { #1 }
  \tl_if_empty:NTF \l_miguel_label_tl
   {
    \miguel_includegraphics:Vn \l_miguel_label_clist { #2 }
   }
   {
    \SetHorizontalCoffin\imagecoffin
     {
      \miguel_includegraphics:Vn \l_miguel_label_clist { #2 }
     }
    \SetHorizontalCoffin\labelcoffin
     {
      \raisebox{\depth}
       {
        \bool_if:NTF \l_miguel_label_box_bool
         { \fcolorbox{white}{white}{\l_miguel_label_size_tl\l_miguel_label_tl} }
         { \l_miguel_label_size_tl\l_miguel_label_tl }
       }
     }
    \SetVerticalPole\imagecoffin{left}{3pt+\CoffinWidth\labelcoffin/2}
    \SetVerticalPole\imagecoffin{right}{\Width-3pt-\CoffinWidth\labelcoffin/2}
    \SetHorizontalPole\imagecoffin{up}{\Height-3pt-\CoffinHeight\labelcoffin/2}
    \SetHorizontalPole\imagecoffin{down}{3pt+\CoffinHeight\labelcoffin/2}
    \use:x{\JoinCoffins\imagecoffin[\l_miguel_label_pos_tl]\labelcoffin[vc,hc]}
    \TypesetCoffin\imagecoffin
   }
   \group_end:
 }
\NewDocumentCommand{\setlabel}{m}
 {
  \keys_set:nn { miguel/label } { #1 }
 }

\cs_new_protected:Nn \miguel_includegraphics:nn
 {
  \includegraphics[#1]{#2}
 }
\cs_generate_variant:Nn \miguel_includegraphics:nn { V }

\ExplSyntaxOff

\makeatletter
\newcommand{\twocolumncaption}{\@dblarg\@twocolumncaption}
\def\@twocolumncaption[#1]#2{%
  \renewcommand{\@makecaption}[2]{%
    \par\vskip\abovecaptionskip\begingroup\small\rmfamily
    \splittopskip=0pt
    \setbox\@tempboxa=\vbox{
      \@arrayparboxrestore \let \\\@normalcr
      \hsize=.5\hsize \advance\hsize-1em
      \let\\\heading@cr
      \noindent ##1\ ##2\par% this line for aastex
    }%
    \vbadness=10000
    \setbox\z@=\vsplit\@tempboxa to .55\ht\@tempboxa
    \setbox\z@=\vtop{\hrule height 0pt \unvbox\z@}
    \setbox\tw@=\vtop{\hrule height 0pt \unvbox\@tempboxa}
    \noindent\box\z@\hfill\box\tw@\par
    \endgroup\vskip \belowcaptionskip
  }%
  \setlength{\abovecaptionskip}{4ex}%
  \caption[#1]{#2}%
}
\usepackage{graphicx}
\begin{document}

\title{Material-Driven Optimization of Transmon Qubits for Scalable and Efficient Quantum Architectures}
\author{Jonnalagadda Gayatri}
\author{S.Saravana Veni}
\email{s\textunderscore saravanaveni@cb.amrita.edu}
\affiliation{Department of Physics, Amrita Vishwa Vidyapeetham, Coimbatore, Tamil Nadu, India
}

\begin{abstract}
One of the most crucial steps in creating practical quantum computers is designing scalable and efficient superconducting qubits. Coherence times, connections between individual qubits, and reduction of environmental noise are critical factors in the success of these qubits. Because they can be lithographically fabricated and are less sensitive to charge noise, superconducting qubits, especially those based on the Transmon architecture, have emerged as top contenders for scalable platforms. In this work, we use a combination of design iteration, material analysis, and simulation to tackle the superconducting qubit optimization challenge. We created transmon-based layouts for 4-qubits and 8-qubits using Qiskit Metal and conducted an individual analysis for each qubit. We investigated anharmonicity and extracted eigenfrequencies, computing participation ratios across several design passes, and identifying the top five energy eigenstates using Ansys HFSS. We then created a 2-D cross-section of a single qubit design in COMSOL Multiphysics to evaluate how different materials affect performance. This enables us to assign various superconducting materials and substrates and investigate their effects on energy loss and electromagnetic properties. Qubit coherence and overall device quality are significantly influenced by the materials chosen. This integrated framework of material-based simulation and circuit design offers a workable way to create reliable superconducting qubit systems and supports continued attempts to create scalable, fault-tolerant quantum computing. 
\end{abstract}

\maketitle

\section{Introduction}

Quantum computing stands at the forefront of technological innovation, offering novel approaches to solving computational problems that are intractable for classical systems. Its uses cover a wide range of fields, such as complicated optimization issues, quantum many-body simulations, and developments in cryptography protocols. Superconducting qubits have become a top contender among the many platforms created for quantum computation because of their incredibly fast gate speeds, compatibility with well-established semiconductor fabrication methods, and scalable integration with current microwave and cryogenic technologies. 

Superconducting qubits operate based on the quantum mechanical properties of superconductors—materials that exhibit zero electrical resistance at cryogenic temperatures. These qubits are typically realized using Josephson junctions, which allow for the controlled tunneling of Cooper pairs and enable the formation of discrete energy levels required for quantum logic operations \cite{ref1}. The development of superconducting qubit designs has progressed rapidly since the introduction of the Cooper-pair box in the late 1990s. Subsequent architectures such as the Transmon and Fluxonium were engineered to mitigate sensitivity to noise sources and extend coherence times. The Transmon design, in particular, has gained wide popularity due to its reduced susceptibility to charge noise and its seamless integration into scalable quantum systems \cite{ref2}.

Building large-scale, fault-tolerant quantum processors is still a challenge despite tremendous progress.  A thorough understanding of the interactions between the qubit shape, material characteristics, electromagnetic behavior, and ambient factors is necessary to accomplish this goal.  Optimizing qubit performance and coherence requires a comprehensive strategy that combines systematic simulations, iterative hardware improvements, and quantum device modeling.  To guide design decisions in this situation, it becomes essential to analyze qubit properties including energy level structures, anharmonicity, and field participation across various materials and layouts.

In this research, we investigate superconducting qubit systems based on the Transmon architecture using a combination of Qiskit Metal and COMSOL Multiphysics. Qiskit Metal is an open source framework for the design of superconducting quantum chips and devices. COMSOL Multiphysics, on the other hand, provides a platform to simulate microwave structures and analyze electromagnetic properties. Our study begins with the design and simulation of two multi-qubit configurations: a 4-qubit and an 8-qubit chip. These systems were evaluated to understand the impact of scaling on important quantum properties. Individual qubits within each chip were analyzed to extract their resonance frequencies, calculate the eigenenergies of the first five quantum states, and assess anharmonicity. This comparison provides insight into the working of these qubits.

To deepen the material-level analysis, we extended our study by performing a two-dimensional cross-sectional simulation of a single Transmon qubit using COMSOL. This analysis was performed for two different sets of superconducting materials and substrates, namely, combinations such as aluminum and niobium on a silicon substrate.

This research presents a multilayered approach to understanding the performance of superconducting qubits, ranging from material-driven coherence enhancements to architectural scaling. The 2D material-specific work offers crucial information on how substrate and superconductor selections impact quantum device performance, while our comparison of 4- and 8-qubit layouts clarifies how circuit complexity affects qubit behavior. System- and material-level simulations work together to create a solid foundation for upcoming experiments and the construction of large-scale quantum processors.  This research adds to ongoing attempts to create scalable, dependable, and high-performance superconducting quantum computers by fusing design tools with simulations based on physics \cite{ref3}.

\section{METHODOLOGY}
In this study, we used a combined simulation workflow involving Qiskit Metal, Ansys HFSS, and COMSOL Multiphysics to design and analyze superconducting transmon qubits. The goal was to explore both scalable qubit architectures and material-dependent performance characteristics through multiplatform simulations.

We began by designing two chip layouts, one containing four qubits and another with eight, using Qiskit Metal. These layouts included transmon qubits coupled to readout resonators and inter-qubit coupling structures. Parameterized components in the design allowed us to test variations in layout geometry, which are critical in scalable quantum systems. Qiskit Metal’s modular environment made it convenient to construct and export designs for further high-frequency simulations \cite{ref4, ref5}.

The geometries were exported from Qiskit Metal to Ansys HFSS, a widely used finite element solver for electromagnetic field simulations in quantum circuits. Within HFSS, we performed eigenmode analyses to extract the resonance frequencies of each qubit. Simulations were conducted across multiple convergence passes to ensure high precision. From the simulated spectra, we calculated the eigenenergies of the first five quantum levels of each qubit and plotted their anharmonicities, which are essential for determining how distinguishable qubit states are from each other. We also plotted the wavefunctions of individual qubits. This level of electromagnetic analysis is consistent with previous work in the field that relies on HFSS for optimizing quantum hardware components \cite{ref6}.

We manually built a two-dimensional cross-sectional model of a single transmon qubit in COMSOL Multiphysics to examine the effects of various materials on qubit performance. This method kept processing demands low while enabling high-resolution characterization of electromagnetic activity close to material interfaces. Because of its great geometric precision in simulating dielectric involvement and field confinement, COMSOL was selected. To mimic boundary conditions and precise structural details close to the capacitor pads and Josephson junction, an airbox and a user-controlled mesh were included.

We simulated two configurations: both used a silicon substrate, but one employed aluminum as the superconducting metal and the other used niobium. These materials are commonly used in superconducting quantum devices due to their established fabrication compatibility and differing coherence characteristics. Niobium is known for higher critical temperature and mechanical stability, while aluminum-based devices have demonstrated long coherence times, particularly in planar transmon designs \cite{ref7,ref8}. To model the electric field behavior realistically, we grounded one of the metal pads and applied a potential to the other, thereby creating an electric potential difference across the structure. This setup emulates the charging and field conditions present in a transmon qubit and enables the extraction of material-dependent electric field and energy density distributions. The 2D cross-sectional model allowed us to isolate these electromagnetic effects at the material level without interference from full 3D or system-level complexities.

Our Ansys HFSS 4-qubit and 8-qubit simulations are enhanced by this material-focused research. While the 8-qubit arrangement revealed layout-induced problems including frequency congestion and crosstalk, the 4-qubit setup provided a baseline for isolated qubit properties.  When combined, these techniques offer a thorough framework for maximizing the performance of superconducting qubits from an architectural and material standpoint.

\section{RESULTS AND DISCUSSION}
In order to determine the electromagnetic properties and performance of our designs, we used Ansys HFSS to generate 3D models for both 4-qubit and 8-qubit structures. These qubits are based on Transmon architecture, which are composed of readout resonators and a non-linear Josephson Junction. The configurations were designed to preserve qubit isolation while enabling the possibility of scalable inter connectivity (see Fig. 1). Important characteristics such as coupling, and resonant frequencies were extracted using these models.

\begin{figure}
    \centering
    \includegraphics[width=1.0\linewidth]{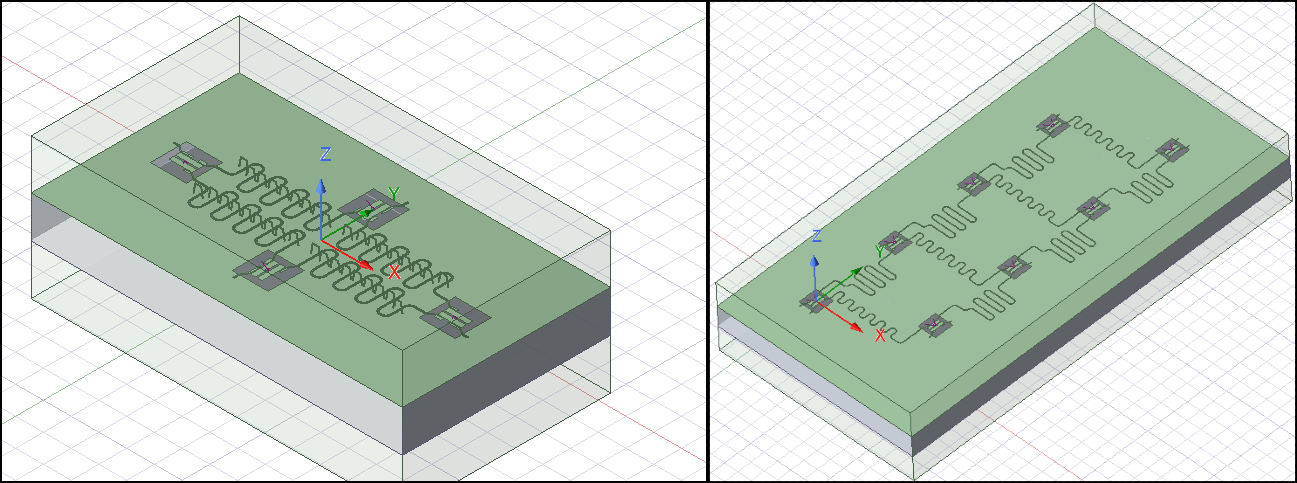}
    \caption{3D models of superconducting qubit chips generated using Ansys HFSS. The left panel shows a 4-qubit transmon layout, while the right panel displays an 8-qubit transmon configuration. Each qubit consists of a Josephson junction-based circuit with coupled resonators, patterned on a substrate and enclosed within a simulation domain for electromagnetic analysis. These models are used to study qubit interactions}
    \label{Figure 2}
\end{figure}

The capacitance and inductance values were specified during the design stage. These values directly impact the resonant and anharmonicties of the transmon qubit and ensure that they operate within the intended range. 

\begin{figure}
    \centering
    \includegraphics[width=1.0\linewidth]{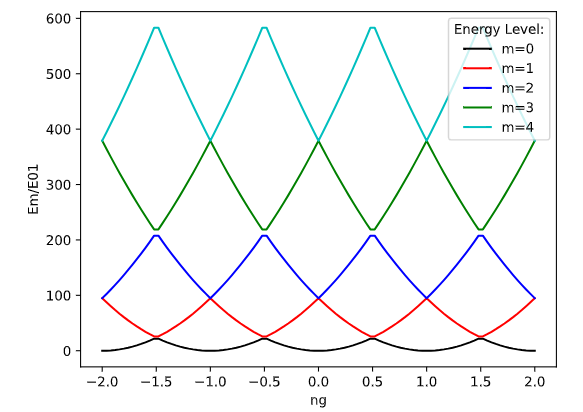}
    \caption{Eigenenergies Em of the first five levels (m = 0, 1, 2, 3, 4) of the transmon Hamiltonian as a 
function of the effective offset charge ng for various ratios of EJ /EC . Energies are normalized to the
 transition energy E01 at the charge degeneracy point (ng = 0.5). The zero-point of energy is taken at the 
bottom of the m = 0 level.}
    \label{fig:enter-label}
\end{figure}

Fig. 2 illustrates the eigenenergies of the first five levels ($m = 0$ to $4$) of the transmon Hamiltonian as a function of the effective offset charge $n_g$, with energies normalized to the transition energy $E_{01}$ at the charge degeneracy point ($n_g = 0.5$). The transmon is governed by the Hamiltonian
\[
\hat{H} = 4E_C(\hat{n} - n_g)^2 - E_J \cos \hat{\phi}
\]
where $E_C$ is the charging energy, $E_J$ is the Josephson energy, $\hat{n}$ is the number operator for Cooper pairs, and $\hat{\phi}$ is the superconducting phase difference across the junction. The figure shows that the energy bands become flatter with increasing $E_J/E_C$, especially for lower levels, indicating strong suppression of charge dispersion. This flattening reflects the transmon’s insensitivity to charge noise—a key advantage over the Cooper Pair Box. The periodicity in $n_g$ stems from the $2e$-periodicity in the charge basis, and the energy reference is set so that the bottom of the $m = 0$ level corresponds to zero.

\begin{figure}
\centering
\includegraphics[width=\linewidth]{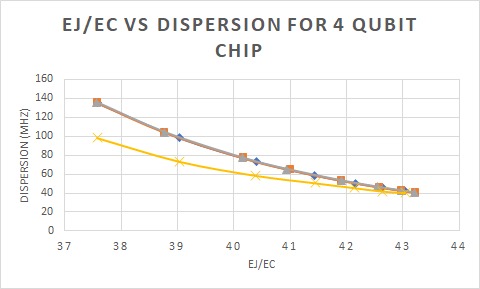}
\small Electric field distribution (a) 
\includegraphics[width=\linewidth]{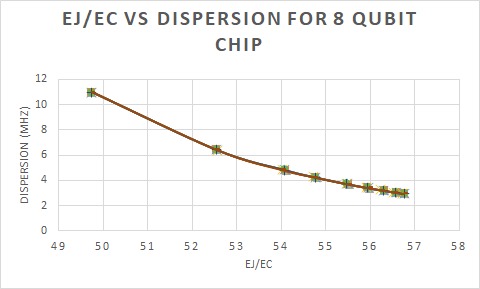}
\small (b)
\caption{The graphs presents the relationship between the ratio of Josephson energy to charging energy (EJ/EC) and the corresponding dispersion values for four and eight transmon qubits on a superconducting chip.}
\end{figure}

Fig. 3 describes the dispersion versus $E_J/E_C$ plots for the 4-qubit and 8-qubit transmon chips. It reveals that dispersion decreases as the $E_J/E_C$ ratio increases, consistent with the transmon regime where larger $E_J$ suppresses sensitivity to charge noise. While the 4-qubit chip shows higher dispersion values (20–160~MHz) and greater variability among qubits, the 8-qubit chip exhibits significantly lower dispersion (2–12~MHz) with tightly clustered data. This suggests improved fabrication uniformity, better environmental isolation, and enhanced stability in the 8-qubit design—key attributes for reliable multi-qubit operation and scalability in quantum processors.

\begin{figure}
\centering
\includegraphics[width=\linewidth]{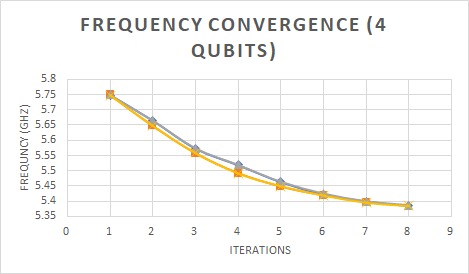}
\small (a) 
\vspace{0.2cm}  % Optional: Adds space between the two images
\includegraphics[width=\linewidth]{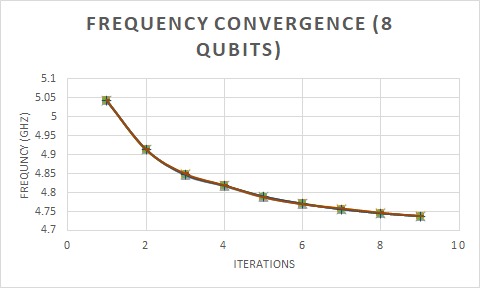}
\small (b)
\caption{Convergence of qubit resonance frequencies over simulation passes (iterations) for superconducting transmon qubit arrays. The top panel shows results for the 4-qubit design, and the bottom panel corresponds to the 8-qubit design. Each colored line represents the real part of the resonant frequency of an individual qubit.}
\end{figure}

The frequency convergence plots for the 4-qubit and 8-qubit systems demonstrate how qubit frequencies stabilize over successive simulation or calibration iterations (see fig. 4). The 4-qubit system shows slower and less uniform convergence, with noticeable variation among qubits, indicating possible design inconsistencies or numerical sensitivity. In contrast, the 8-qubit system converges more rapidly and uniformly to a stable frequency, reflecting a more robust design and improved parameter uniformity. These results suggest that the 8-qubit chip offers greater simulation reliability and design stability, both critical for scalable quantum device performance.

\begin{figure} 
\centering
\includegraphics[width=\linewidth]{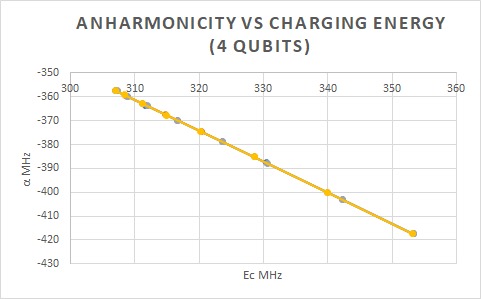}
\small (a)
\vspace{0.2cm}  % Optional: Adds space between the two images
\includegraphics[width=\linewidth]{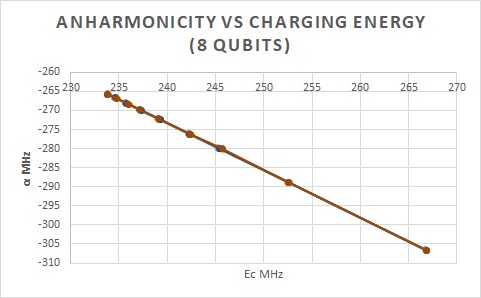}
\small (b)
\caption{Comparison of anharmonicity versus charging energy ($E_C$) for 4-qubit and 8-qubit transmon chips. In both cases, the anharmonicity decreases linearly with increasing $E_C$, consistent with the theoretical relation $\alpha \approx -E_C$.}
\end{figure}

The relationship between anharmonicity ($\alpha$) and charging energy ($E_C$) in both the 4-qubit and 8-qubit transmon chips, given by fid. 5, shows a consistent linear trend with a negative slope, reflecting the theoretical expectation $\alpha \approx -E_C$ derived from the transmon model. This relation arises because transmon qubits, designed to reduce charge noise by operating in the regime $E_J \gg E_C$, exhibit weakly anharmonic energy levels where the spacing between $\lvert 1 \rangle$ and $\lvert 2 \rangle$ states decreases with increasing $E_C$. 
The 8-qubit chip exhibits tighter clustering of data points around the fit line, suggesting more homogeneity in qubit parameters, though both chips follow this pattern.  This implies better design or fabrication consistency, which makes control, calibration, and error mitigation easier throughout the processor and is crucial for scalable quantum computing.

\begin{figure}
\centering
\includegraphics[width=\linewidth]{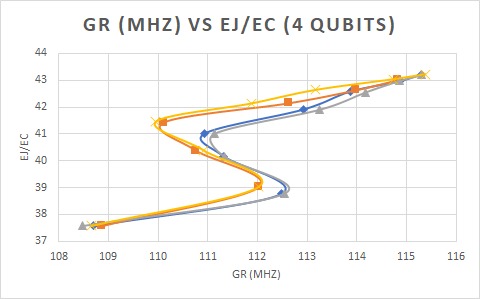}
\small (a) 
\vspace{0.2cm}  % Optional: Adds space between the two images
\includegraphics[width=\linewidth]{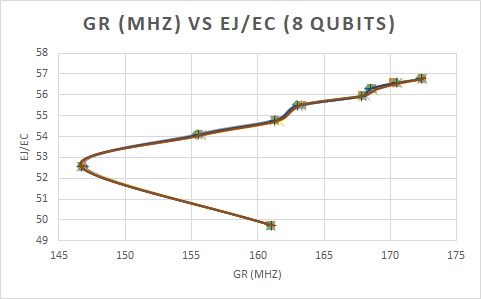}
\small (b) 
\caption{Non-monotonic relationship between qubit-resonator coupling strength (gr) and $E_C/E_J$ ratio, illustrating the trade-offs between readout speed, coherence, and system size in multi-qubit superconducting qubit systems.}
\end{figure}

To understand the behaviour of qubit--resonator coupling, we analyze the variation of the coupling strength $g_r$ as a function of the transmon parameter ratio $E_J/E_C$, where $E_J$ is held constant and $E_C$ is varied (see fig. 6). This approach reflects how the coupling strength changes as the qubit frequency, $f_q \propto \sqrt{8E_J E_C} - E_C$, is swept relative to the fixed resonator frequency. The resulting plots for different multi-qubit systems reveal characteristic non-monotonic behaviour in $g_r$, with observed S-shaped or U-shaped dependencies on $E_J/E_C$. These shapes emerge from the complex interplay between the transmon frequency and its coupling to the resonator mode, highlighting that $g_r$ does not vary linearly with transmon parameters. Fabrication uniformity is reflected in the dispersion or clustering of these curves across various qubits on the same chip; tighter clustering denotes more consistent device parameters. Because they direct the choice of operating regimes that strike a compromise between strong coupling, advantageous anharmonicity, and decreased charge sensitivity, these observations are crucial for qubit design and calibration.

\begin{figure}
\centering
\includegraphics[width=\linewidth]{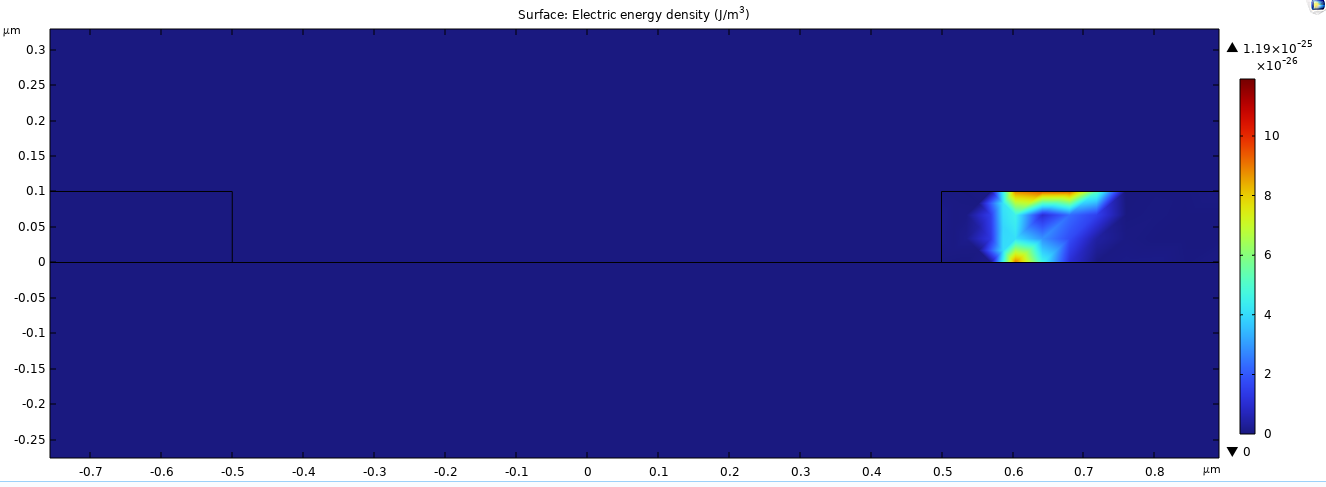}
\small (a)
\vspace{0.2cm}  % Optional: Adds space between the two images
\includegraphics[width=\linewidth]{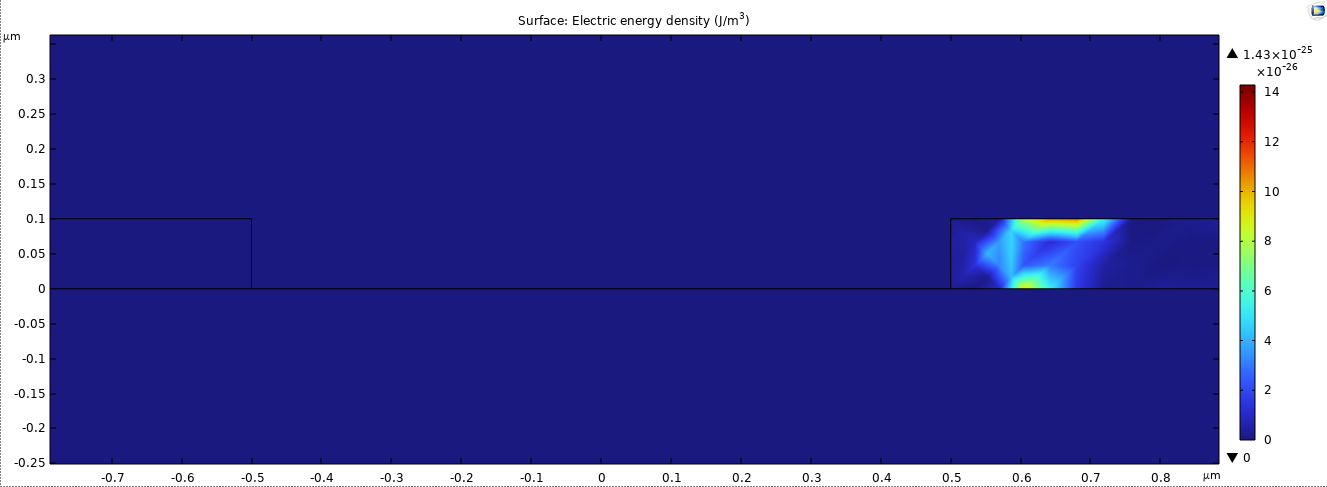}
\small (b)
\caption{Electric energy density distribution in microstructures with Aluminum (a) and Niobium (b) superconductors on a Silicon substrate.}
\label{}
\end{figure}
%%%%%%%%%%%%%%%%%%%%%%%%%%%%%%%%%%%%%%%%%%%%%%%%%%%%
\begin{figure}
\centering
\includegraphics[width=\linewidth]{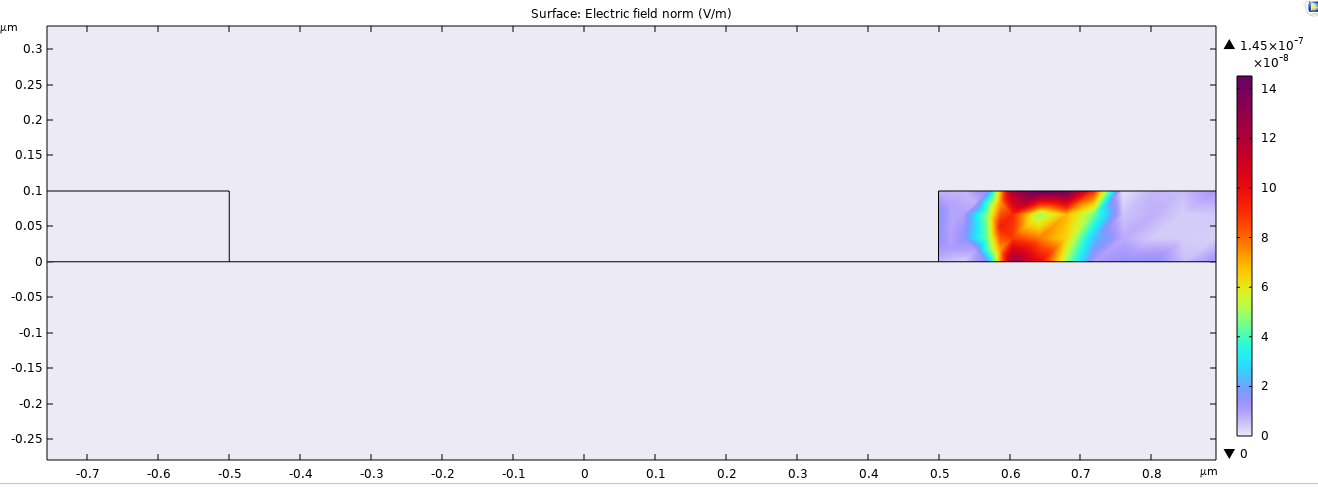}
\small (a)
\vspace{0.5cm}
\includegraphics[width=\linewidth]{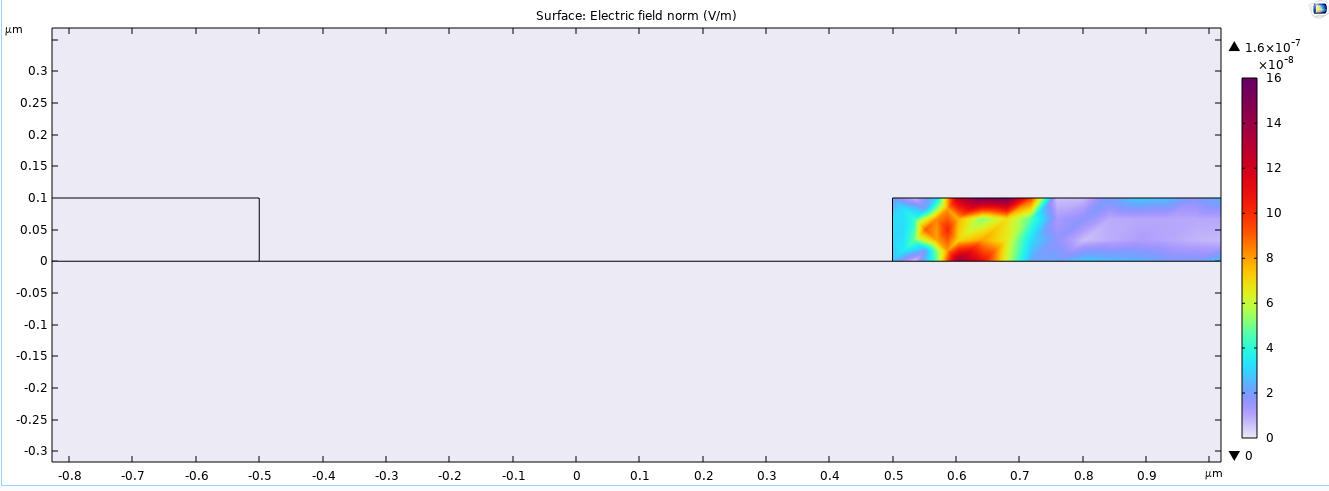}
\small (b) 
\caption{Electric field magnitude in superconducting microstructures with Aluminum (a) and Niobium (b) superconductors on a Silicon substrate.}
\end{figure}
%%%%%%%%%%%%%%%%%%%%%%%%%%%%%%%%%%%%%%%%%%%%%%%%%%%%%%%%%%%
Fig. 7 and 8 compare the electric field norm and electric energy density distributions for Aluminum (Al) and Niobium (Nb) superconducting configurations on a silicon substrate, revealing important material-dependent differences relevant to qubit performance. Al confines the electric field more firmly to the metal edges in both figures, whereas the Nb-based arrangement shows wider and more diffuse electric field distributions. This discrepancy is primarily attributed to Nb's larger kinetic inductance and greater London penetration depth, which cause the electromagnetic field to penetrate deeper into the substrate and surrounding dielectrics. Higher peak electric field magnitudes are observed near critical surfaces due to stronger field localization at the metal-dielectric interface, caused by Al's lower kinetic inductance and the formation of a native Al$_2$O$_3$ oxide layer. 

The electric energy density, which is proportional to the electric field strength and the permittivity of the material, is directly impacted by these effects. Whereas Al concentrates more energy in confined, lossy dielectric areas, Nb tends to disperse stored electric energy over a greater volume, lowering peak energy density near interfaces. It is well known that this kind of localization increases the participation of surface two-level systems (TLS), which limits coherence times and significantly contributes to dielectric losses. Furthermore, less abrupt field localization occurs in Nb structures due to the smaller dielectric contrast between Nb and silicon (as opposed to Al and silicon), which further reduces energy concentration at defect-prone boundaries.

Overall, the simulation results indicate that niobium provides a more favourable electromagnetic environment for superconducting qubits by reducing dielectric participation, lowering surface losses, and improving coherence. These observations align with the broader understanding of superconducting material behaviour at microwave frequencies and underscore the importance of material choice in optimizing qubit design for noise resilience and scalability.

\section{Conclusions}

In this work, we used finite element simulations to perform a comparative electromagnetic analysis of superconducting qubit designs based on various materials and geometries. In particular, we investigated the electric field norm and energy density distributions for aluminum (Al) and niobium (Nb) superconducting layers on a silicon substrate, as well as the frequency convergence behavior of 4-qubit and 8-qubit systems.

Our findings demonstrate that the 8-qubit design exhibits better frequency convergence than the 4-qubit system, showing faster, smoother, and more consistent behavior. This suggests improved simulation stability and likely enhanced physical robustness. These results imply that the 8-qubit architecture benefits from enhanced layout symmetry, coupling homogeneity, or reduced susceptibility to numerical instabilities.

In terms of material, we found that Al-based structures exhibit better field localization at metal-dielectric interfaces, whereas Nb-based designs exhibit broader and more dispersed electric fields and energy densities.  These variations stem from the materials' inherent electromagnetic characteristics; Nb has a higher kinetic inductance, deeper field penetration, and more advantageous oxide behavior, which all help to lower energy concentration in lossy regions and reduce dielectric participation.  Al, on the other hand, is more vulnerable to surface-related decoherence mechanisms due to its severe field confinement.

When combined, our results highlight the significance of material choice and circuit architecture in the design of high-performance superconducting qubit systems.  When implemented with Nb instead of Al, the qubit architecture shows promising properties in terms of electromagnetic field behavior and simulation convergence.  These discoveries lay the groundwork for developing next-generation superconducting qubit devices that are more scalable and coherent for use in quantum computing and sensing.

\section*{Acknowledgements}

The Acknowledgements section is not numbered. Here you can thank helpful
colleagues, acknowledge funding agencies, telescopes and facilities used etc.

\vfill

\end{document}